\documentclass[%
 reprint,
 amsmath,amssymb,
 aps,
]{revtex4-2}

\usepackage[dvipdfmx]{graphicx}% Include figure files
\usepackage{dcolumn}% Align table columns on decimal point
\usepackage{bm}% bold math
\usepackage{color}
\begin{document}

\preprint{APS/123-QED}

\title{Effect of mobility on synchronization of nonlocally coupled oscillators with a phase lag}

\author{Bojun Li}
\author{Nariya Uchida}
\email{uchida@cmpt.phys.tohoku.ac.jp}
\affiliation{Department of Physics, Tohoku University, Sendai 980-8578, Japan}

\date{\today}% It is always \today, today,

\begin{abstract}
Nonlocally coupled oscillators with a phase lag self-organize into various patterns
such as global synchronization, the twisted state, and the chimera state.
In this paper, we consider nonlocally coupled oscillators that move on a ring
by randomly exchanging their positions with the neighbors, and
investigate the combined effects of phase lag and mobility on the collective phase dynamics.
Spanning the whole range of phase lag and mobility, we show that
mobility promotes synchronization for an attractive coupling,
while it destroys coherence for a repulsive coupling.
The transition behaviors are discussed in terms of
the timescales of synchronization and diffusion of the oscillators.
We also find a novel spatiotemporal pattern at the border between
coherent and incoherent states.
\end{abstract}

\maketitle

%\tableofcontents

%%%%%%%%%%%%%%%%%%%%%%%%%%%%%%%%%%%%%%%%%%%%%%%%%%%%%%%%%%%%%%%%%%%%%%%%%%%%%%%%%%

\section{Introduction}
Collective synchronization
is observed in a wide range of natural phenomena,
such as flashing fireflies, cardiac beats,
and circadian rhythms~\cite{rosenblum2003synchronization}.
While a global coupling explains the emergence of fully synchronized
states~\cite{acebron2005kuramoto},
a nonlocal (or finite-range) coupling introduces 
the twisted and the chimera states.
The twisted state~\cite{wiley2006size,girnyk2012multistability} is a state in which
the phase difference between two neighbor oscillators is constant.
The chimera state~\cite{Panaggio_2015,parastesh2021chimeras} is characterized by
coexistence of coherent and incoherent groups of oscillators,
which was discovered
by introducing a phase lag in the nonlocal coupling~\cite{kuramoto2002}.
These states are most simply described by the phase oscillators on a linear array,
the time evolution of which obeys
\begin{equation}
\dot{\phi}(x,t) = \omega_0 - \sum_{x' \neq x} g(x - x')
\sin \left[ \phi(x,t) - \phi(x',t) + \alpha \pi \right],
\label{eq.main}
\end{equation}
where $\phi(x,t)$ is the phase at position $x$ and time $t$,
$\omega_0$ is the intrinsic phase velocity, $g(x)$ is the interaction kernel,
and $\alpha \pi$ is the phase lag
(we assume $0 \le \alpha \le 1$ without loss of generality).
The phase lag induces frustration in the system
which gives rise to a variety of intriguing patterns.
For $\alpha<0.5$, the coupling is attractive.
If $\alpha$ is sufficiently small,
the system that started from a random initial condition eventually
reaches a coherent state (either a synchronous or a twisted state).
When $\alpha$ exceeds a threshold, the frustration destroys coherence
and forms a chimera state or
a multichimera state with many coherent and incoherent regions~\cite{maistrenko2014cascades}.
The onset of the multichimera state is a critical transition
that resembles directed percolation~\cite{duguet2019loss,kawase2019}.
For a repulsive coupling ($\alpha > 0.5$),
we find coexistence of twisted states
with positive and negative phase differences
separated by incoherent strips~\cite{li2021large}.
A randomly branching multichimera state disappears at
an upper threshold of $\alpha$ and is replaced by
stripe patterns with smooth boundaries.
For $\alpha$ close to unity, frequency synchronization
takes place in the coexistent twisted states.

Another interesting situation is brought by
the oscillators that move in space and change the partners to interact with.
Mobile oscillators are related to biological phenomena including animal groups~\cite{animals} and chemotactic elements~\cite{chemotaxis},
and also have potential technological applications,
i.e., in robotics~ \cite{robotics,robotics2} and
wireless sensor networks~\cite{wireless-sensor}.
The effect of mobility on synchronization has been an actively studied subject in the last decade~\cite{peruani2010mobility,fujiwara2011synchronization,uriu2013dynamics,gomez2013motion,buscarino2016interaction,fujiwara2016synchronization,majhi2017amplitude,petrungaro2017mobility,o2017oscillators,wang2019chimera,petrungaro2019synchronization,smirnov2021disorder}.
In Ref.~\cite{peruani2010mobility}, it was shown that mobility
destabilizes the twisted states and promotes synchronization.
Ref.~\cite{uriu2013dynamics} shows that mobility speeds up synchronization
by extending the effective coupling range.
Ref.~\cite{petrungaro2017mobility} studied the system with a delayed coupling and found that mobility can induce chimera states.
Ref.~\cite{o2017oscillators} considered mobile oscillators with mutual interaction between motion and phase, and found a variety of nontrivial swarming and synchronized patterns.
However,
little is known about the combined effects of phase lag
and mobility on the twisted and chimera states.
Two recent studies considered the system with a phase lag:
Ref.\cite{wang2019chimera} demonstrated the transition between synchronous state
and
the chimera states with one or two coherent regions;
Ref.\cite{smirnov2021disorder} shows that disorder of the oscillators facilitates
stability of chimera.
These papers focus on a system with the phase lag $\alpha$ close to 0.5,
and a study spanning a wide range of $\alpha$ is still lacking.

In this paper, we investigate the combined effects of phase lag and mobility 
on the collective phase dynamics of nonlocally coupled oscillators on a ring.
We span the whole range of phase lag and mobility, and find that
mobility promotes synchronization for an attractive coupling,
while it destroys coherence for a repulsive coupling.
The transition behaviors are discussed in terms of the timescales of
synchronization and diffusion of the oscillators.
For the attractive case, we estimate
the size of the basins of attraction for the synchronous and twisted states,
and characterize the collapse of the twisted states into the synchronous states
quantitatively.
We also find a novel mesh-like pattern that consists of intersecting traveling waves
near the coherent-chimera transition.
For the repulsive case, mobility induces noisy patches of twisted states,
which gradually crosses over to a fully incoherent state
as the mobility is increased.

%%%%%%%%%%%%%%%%%
\begin{figure*}[htbp]
\includegraphics[width=17.2cm]{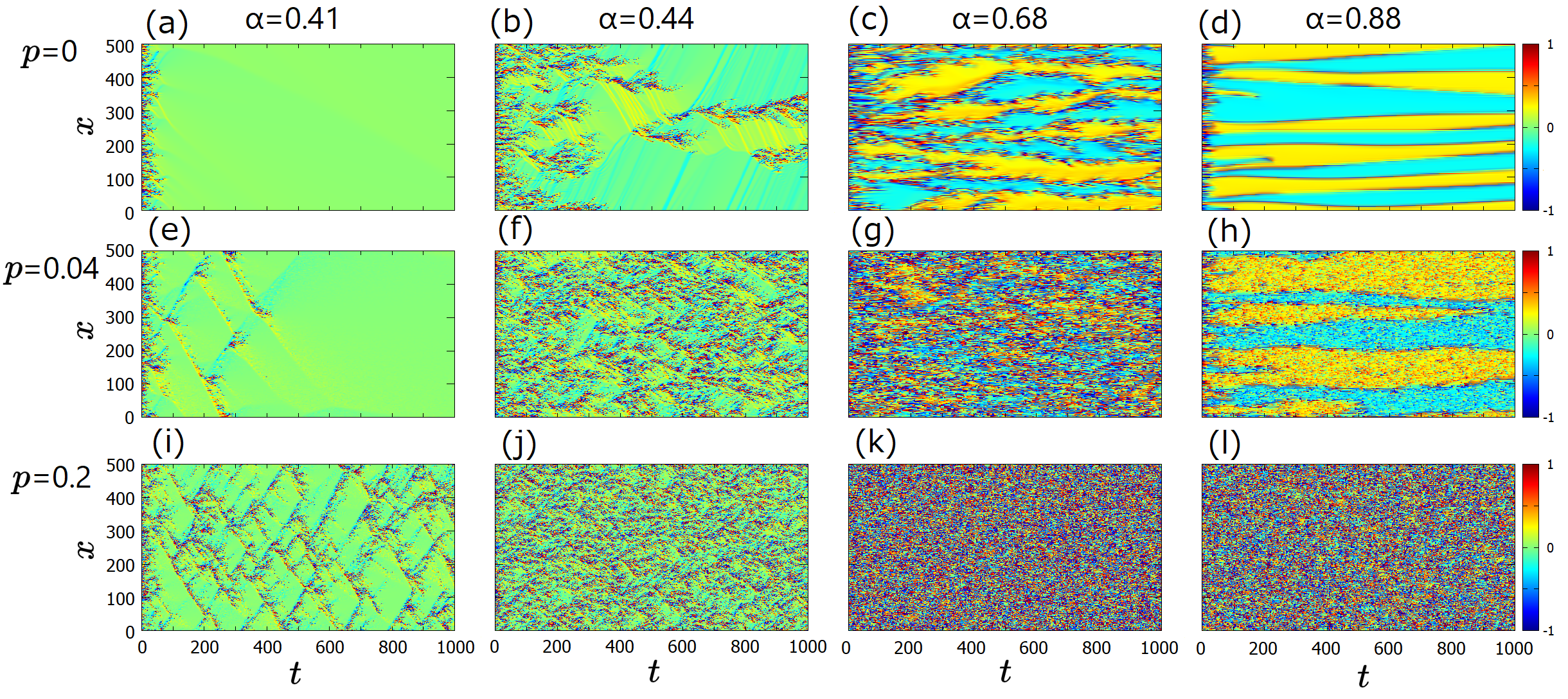}
\caption{(Color online) Spatiotemporal patterns of the phase difference $\Delta(x,t)$, 
for $\alpha=0.41,0.44,0.68,0.88$ from the left to the right column,
and $p=0,0.04,0.2$ from the top to the bottom row.
\label{fig:1}}
\end{figure*}

%%%%%%%%%%%%%%%%%%%%%%%%%%%%%%%%%%%%%%%%%%%%%%%%%%%%%%%%%%%%%%%%%%%%%%%%%%%%%%%%%%

\section{Model}
We consider $N$ oscillators that are distributed on a ring
with $N$ sites
labeled by the integer coordinates $x$ (mod $N$).
Each site on the ring is occupied by a single oscillator
at a time.
The model is built upon Eq.~(\ref{eq.main})
for non-mobile oscillators with the interaction kernel
\begin{equation}
g(x) \left\{
\begin{array}{ll}
\displaystyle
= \frac{1}{2R} & \quad (|x| \le R)
\vspace{3mm}
\\
= 0 & \quad (|x| > R)
\end{array}
\right.
\end{equation}
where $R$ is the coupling range.
The intrinsic frequency $\omega_0$ is set to zero
without losing generality.
The oscillators move stochastically by exchanging their positions 
with their nearest neighbors at a constant rate.
It is implemented by a discrete dynamical rule with a small time step $\delta t$. 
At each time step, 
each oscillator exchanges its position with its right nearest neighbor
with the probability $p\delta t$,
which causes the change of phases
$\phi(x,t+\delta t) =  \phi(x+1,t)$ 
and 
$\phi(x+1,t+\delta t) = \phi(x,t)$.
Since the average number of moves of each oscillator
during one time unit is $2p$,
the mean square displacement (MSD) is given by
\begin{equation}
\langle (\Delta x)^2 \rangle 
\equiv 
\langle \left[ x(t'+t) - x(t') \right]^2 \rangle = 2p t,
\label{eq.MSD}
\end{equation}
and the diffusion constant is $D=p$.
Thus the model is characterized by three parameters $R$, $\alpha$, and $p$. 

We solve Eq.(\ref{eq.main}) using the Runge-Kutta method with a time step $0.01$. Unless otherwise stated, we set $R=5$ and $N=500$.
Uniformly random values of the phases are used for the initial condition.
The mobility is implemented by randomly choosing a site $x$ and exchanging the phases of the oscillators at $x$ and $x+1$
with the probability $p\delta t$. This process is repeated $N$ times
at each time interval $\delta t$, which is also set to $0.01$.
Note that the upper limit of the probability $p\delta t$
is unity and therefore $p\le 100$.
We checked that a smaller value of $\delta t$ does not
significantly change the results presented
in the next section.

\section{Numerical Results}

\subsection{Spatiotemporal patterns and the strength of incoherence}

In Fig.~\ref{fig:1}, we plot the spatiotemporal patterns
of the phase difference between
two neighbor oscillators $\Delta(x,t)=[\phi(x+1,t)-\phi(x,t)]/\pi$,
truncated in the range $[-1: 1)$.
The first row shows the patterns for $p=0$ with $\alpha$ varied.
We varied $\alpha$ in the range $0\le \alpha < 1$,
and only a few samples are shown.
For $\alpha<0.44$, the system reaches a coherent state
with a uniform and very small value of $\Delta$
(Fig.\ref{fig:1}(a)).
For $0.44\le \alpha<0.5$, incoherent regions unceasingly
branch and disappear, forming a multichimera
state~\cite{duguet2019loss,kawase2019}
(Fig.\ref{fig:1}(b)).
For a repulsive coupling ($\alpha>0.5$),
twisted states with positive and negative values of $\Delta$
are separated by incoherent strips of width $\propto R$;
for $\alpha$ not far from $0.5$, 
a randomly branching multichimera state appears (Fig.\ref{fig:1}(c)), which is replaced by
coexisting twisted states with smooth boundaries for a larger $\alpha$ (Fig.\ref{fig:1}(d))~\cite{li2021large}.

The mobility of the oscillators introduces
a novel spatiotemporal pattern.
The patterns for $p=0.04$ and $0.2$ are shown in the second
and third rows of Fig.~\ref{fig:1}, respectively.
For $0.38\le\alpha<0.44$,
we find a mesh-like pattern that consists of intersecting traveling waves (Fig.\ref{fig:1}(e)(i)).
The chimera state for $\alpha = 0.44$ obtains finer structures
of incoherent regions due to the mobility (Fig.\ref{fig:1}(f)(j)).
For $\alpha>0.5$, noisy spots are added to
the multichimera states (Fig.\ref{fig:1}(g)(h)).
At $p=0.2$, the patterns are further randomized and
we cannot recognize the original patterns for $p=0$
(Fig.\ref{fig:1}(k)(l)).

\begin{figure}[htbp]
\includegraphics[width=8.6cm]{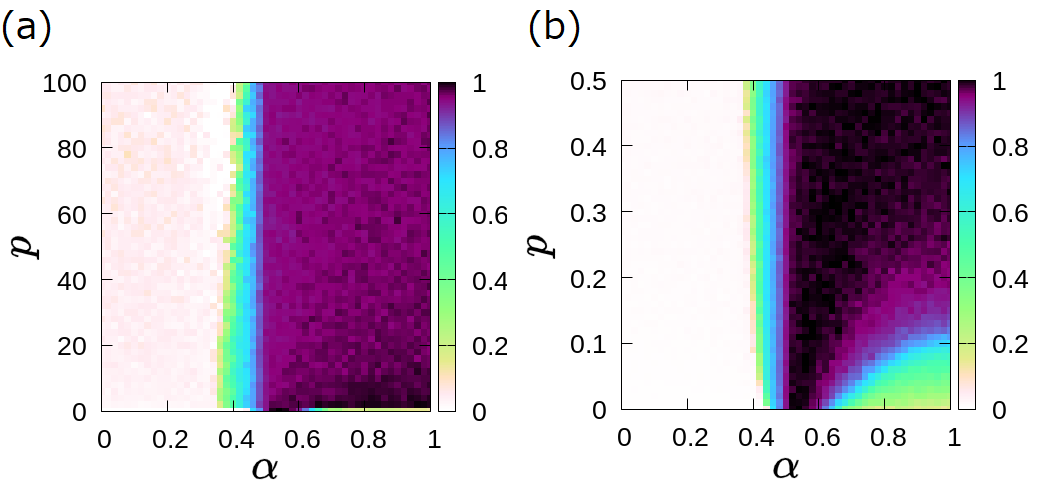}
\caption{
(Color online) The strength of incoherence $\overline{\sigma_\Delta}$
on the $(\alpha,p)$ plane, for
(a) $0\le p< 100$ and (b) $0\le p < 0.5$.
The colorscale (grayscale) shows
a coherent state ($\overline{\sigma_\Delta} \ll 1$) in light (whitish) colors,
and
an incoherent state ($\overline{\sigma_\Delta} \sim 1$) in dark colors (black and dark gray).\label{fig:2}}
\end{figure}

As a measure of the strength of incoherence,
we follow a previous work~\cite{gopal2014observation} 
and
use the standard deviation of
the phase difference within a distance $R$ from the site $x$,
\begin{equation}
\sigma_\Delta(x,t)=\sqrt{\frac{3}{2R}\sum_{x'=x-R}^{x+R-1}[\Delta(x',t)-\bar\Delta(t)]^2},
\label{eq.sigma}
\end{equation}
where the prefactor is chosen to make
the maximum value of $\sigma_\Delta(x,t)$ equal to 1.
The strength of incoherence of the whole system
is written as $\overline{\sigma_\Delta}(t)=\langle \sigma_\Delta(x,t) \rangle_x$.
Fig.~\ref{fig:2} shows the value of $\overline{\sigma_\Delta}(t)$
at $t=1000$ averaged over 10 independent samples for each set of $(\alpha, p)$.
The entire range of the mobility ($0\le p \le 100$) is
shown
in Fig.\ref{fig:2}(a)
and a magnified view of the range $0 \le p \le 0.5$ is
shown
in Fig.\ref{fig:2}(b).
We obtain coherent states ($\overline{\sigma_\Delta} \ll 1$)
for small $\alpha$ irrespective of the value of $p$.
The degree of incoherence sharply increases 
at $\alpha=\alpha_c \simeq 0.4$, and
the threshold $\alpha_c$ has a minor dependence on $p$.
On the other hand, when $\alpha>0.5$, the system falls
into complete incoherence ($\overline{\sigma_\Delta} \sim 1$)
over most of the parameter range
except for very small $p$.
These changes brought by the mobility
will be further discussed in the following sections.
%%%%%%%%%%%%%%%%

\subsection{The effect of mobility on the coherent states}

We first focus on the effect of mobility on the coherent states
for $\alpha < \alpha_c(p)$. 
In this parameter range, the system is multistable, i.e.,
either the synchronous or $q$-twisted states are achieved
depending on the initial condition.
The phase difference in these states can be written as $\Delta=2q/N$, where $q=0$ represents the synchronous state, and
a nonzero integer with $|q|\le\frac{N-1}{2}$ corresponds to
the $q$-twisted state.

\begin{figure}[htbp]
\includegraphics[width=8.6cm]{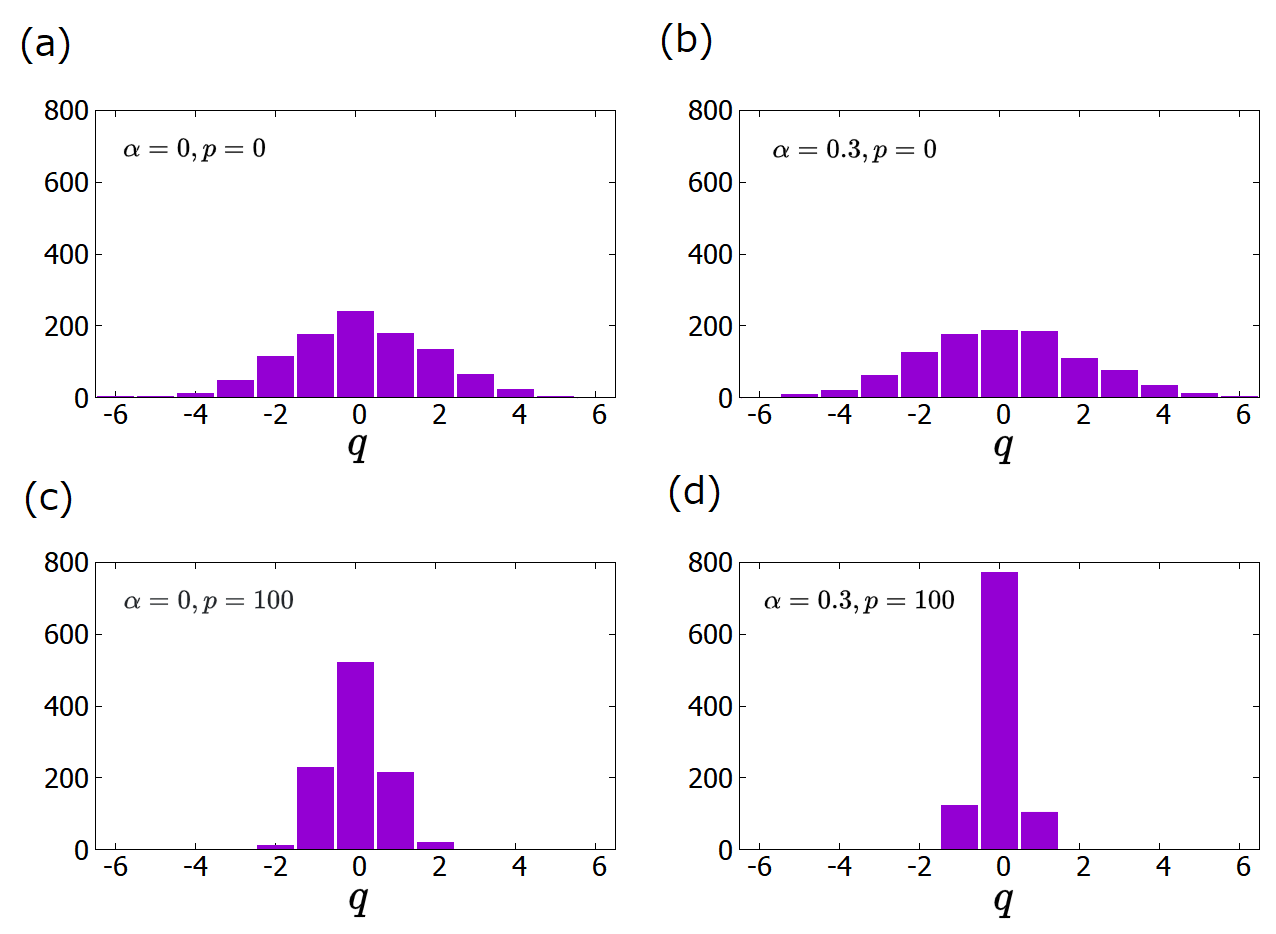}
\caption{Histograms of the winding number $q$ for
1000 independent samples.
(a) $(\alpha,p)=(0,    0)$, which gives $\sigma_q=1.79$,
(b) $(\alpha,p)=(0.3,  0)$, $\sigma_q=2.02$.
(c) $(\alpha,p)=(0,  100)$, $\sigma_q=0.76$.
(d) $(\alpha,p)=(0.3,100)$, $\sigma_q=0.48$.
\label{fig:3}
}
\end{figure}

In the simulation, we compute the winding number $q$ as
\begin{equation}
q=\frac{1}{2}\sum_{x=1}^N\Delta(x, t)
\label{eq:q}
\end{equation}
with $t=1000$.
We take an average over 100 samples 
for each value of $(\alpha, p)$ unless otherwise stated,
including the incoherent states where $\Delta(x,t)$ is not spatially uniform.
We also measure the root mean square (RMS) of $q$, which is denoted by $\sigma_q$.
Fig.~\ref{fig:3} shows the histograms of $q$
for specific values of $(\alpha, p)$.
For $p=0$,
the distribution is wide
and the maximum value of $|q|$ reaches $6$.
As $p$ increases, the distribution becomes narrower,
which means that the mobility enlarges
the size of the basin of attraction for
a smaller value of $|q|$.
Compared at the same mobility $p=100$,
the peak for $\alpha=0.3$ is higher
than the one for $\alpha=0$.
It means that the phase lag widens the basin of attraction
of the synchronous state.

%%%%%%%%%%%%%%%%

\begin{figure}[htbp]
\includegraphics[width=8.6cm]{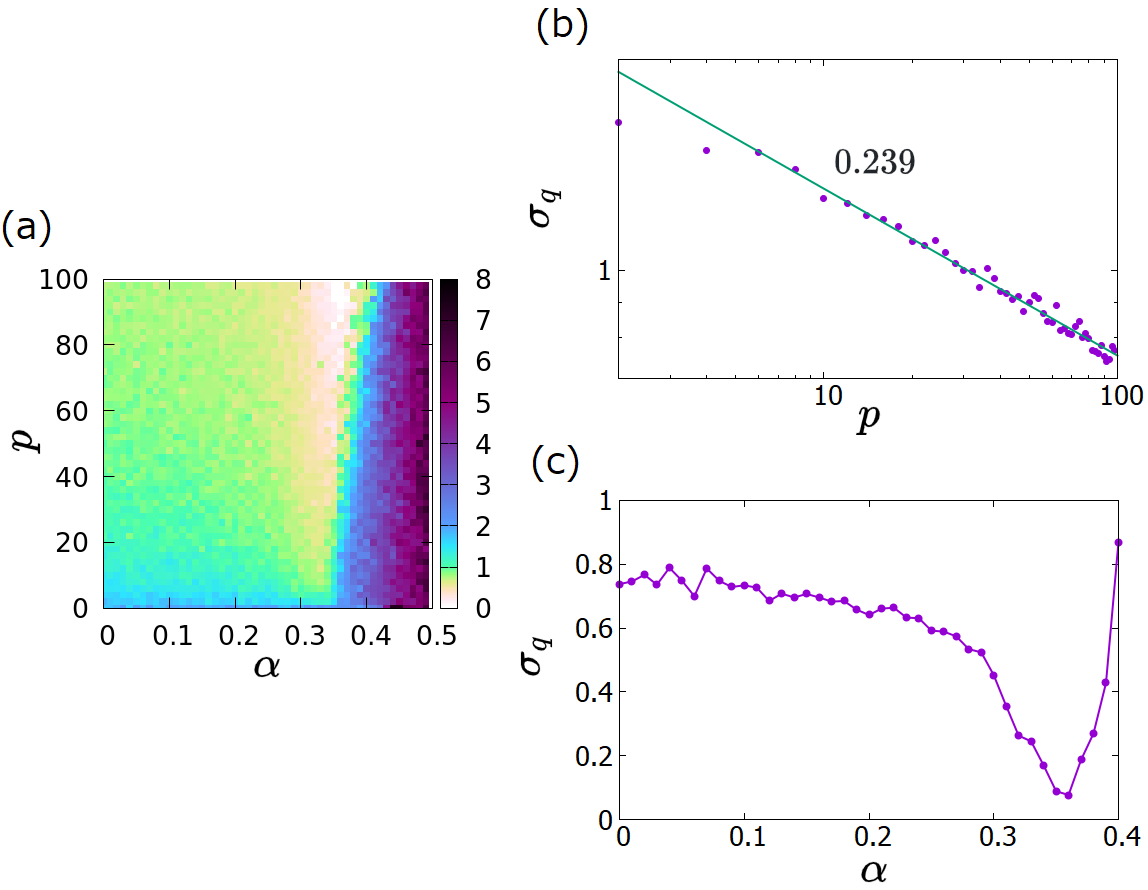}
\caption{(Color online) (a) The RMS of the winding number $\sigma_q$ for
each ($\alpha$, $p$).
(b) The RMS $\sigma_q$ versus $p$ for $\alpha=0$.
(c) The RMS $\sigma_q$ versus $\alpha$ for $p=100$.
The data in (b)(c) are averages over 1000 independent samples.
\label{fig:4}}
\end{figure}

Fig.~\ref{fig:4}(a) shows the RMS
of the winding number on the $(\alpha,p)$ plane.
We focus on the coherent (synchronous or $q$-twisted) states with $\alpha < \alpha_c(p)$ here.
The RMS decreases as $\alpha$ and $p$ increase
in the coherent region of the $(\alpha, p)$ plane.
Notably, it becomes almost zero at large $p$ and near $\alpha=\alpha_c(p)$ (Fig.~\ref{fig:4}(c)),
which means that the synchronous state $q = 0$ is attained for most of the samples.
In Fig.~\ref{fig:4}(b),
we plot $\sigma_q$ for $\alpha=0$ as a function of the mobility parameter $p$.
It is fitted by the power law
$\sigma_q \sim p^{-\beta}$ with the exponent $\beta = 0.239 \pm 0.005$
in the range $10\le p\le 100$. We will discuss in Section IV the reason why the exponent is close to $1/4$.

%%%%%%%%%%%%%%%%

The enhancement of synchronization by the mobility
is explained by disturbance of the $q$-twisted states.
The exchange of two oscillators in a twisted state
alters
the phase differences at three successive sites
from $\{ \Delta,\Delta,\Delta \}$
to  $\{2\Delta,-\Delta,2\Delta\}$.
The total change in the phase differences is $4\Delta$,
and is larger for a larger $|q|$ because $\Delta=2q/N$.
Conversely, a twisted state with a smaller $|q|$ is less
affected, and the fully synchronized state
is unchanged by the exchange.
If the disturbance due to mobility dominates
the effect of interaction to recover the initial (twisted) state,
the system leaves the basin of attraction and hops to another,
as already shown for $\alpha=0$~\cite{wiley2006size}.
In the following, we analyze
the balance between the effects of mobility and coupling
quantitatively, by computing the timescales
to disturb and restore the twisted states.

\begin{figure}[htbp]
\includegraphics[width=8.6cm]{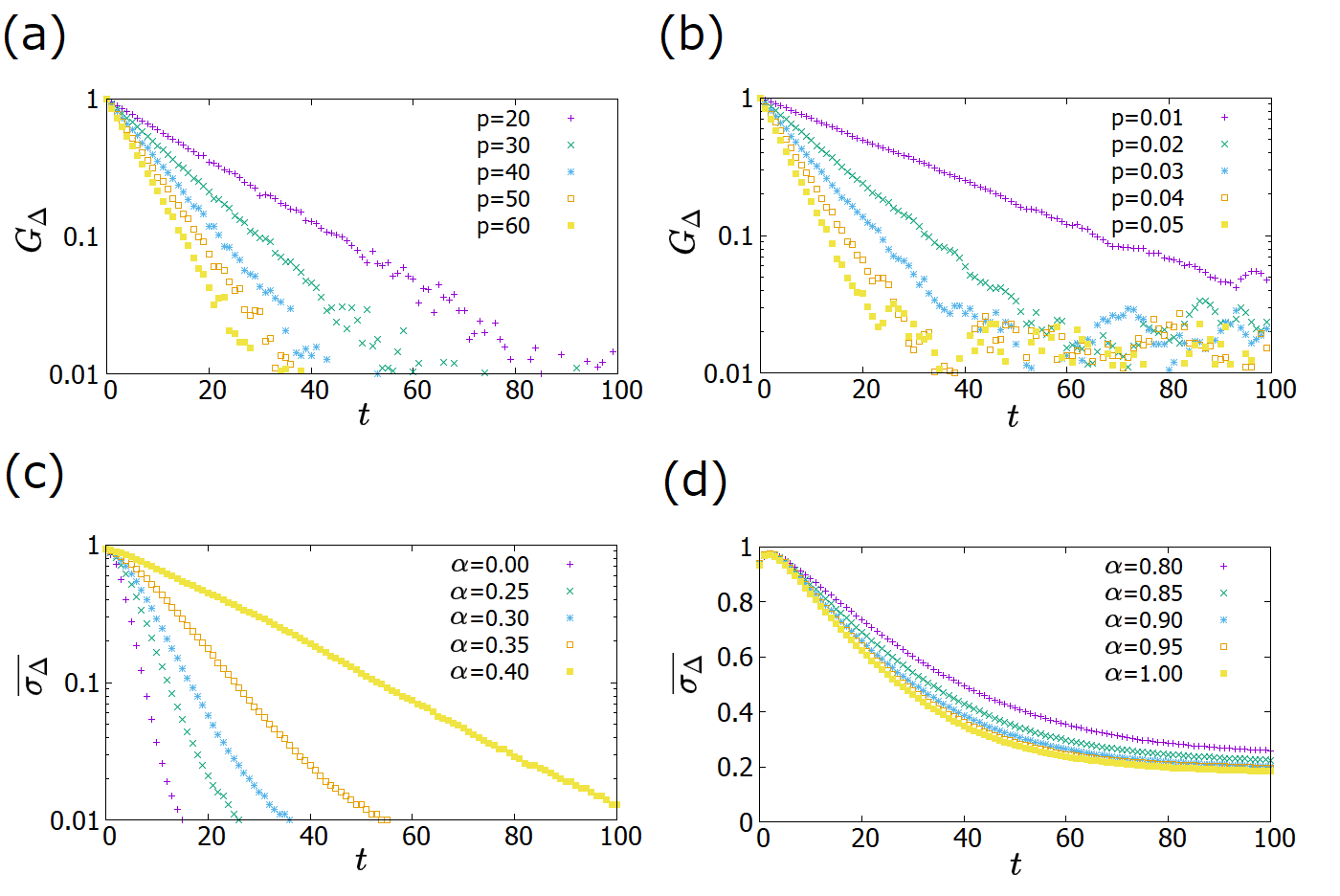}
\caption{(a)(b)
Temporal correlation of the phase difference
$G_\Delta(N/2,t)$, obtained without coupling and
starting from the $q$-twisted state with (a) $q=2$ and (b) $q=67$.
It shows the effect of mobility to randomize the phase difference.
(c)(d) The strength of incoherence $\overline{\sigma_\Delta}$
obtained without mobility and starting from the random state,
for (c) $0\le \alpha\le 0.4$ and (d) $0.8\le \alpha \le 1$.
It shows the effect of coupling to restore the $q$-twisted states.
\label{fig:5}}
\end{figure}

A system in a $q$-twisted state is divided into $|q|$ blocks
of size $N/|q|$. In each block,
the phase changes by $2\pi$.
In the absence of coupling,
the time required for the phase differences to be randomized
should be proportional to the time for each oscillator
to travel the distance of $N/|q|$, the size of a block.
Substituting $N/|q|$ into $\Delta x$ in the left side of Eq.~(\ref{eq.MSD}), the characteristic time for
the randomization is written as
\begin{equation}
\tau_{\rm rand} \sim \frac{N^2}{pq^2}. 
\end{equation}
This can be validated by computing the spatial correlation of phase difference
\begin{equation}
G_\Delta(x,t)
=\left \langle \cos(\Delta(x'+x,t)-\Delta(x',t)) \right \rangle_{x'}
\label{eq.correphi}
\end{equation}
of a system with only the mobility $p$ but no phase coupling.
Here the average is taken over $x'$ and 100 independent samples. 
The function has little dependence on $x$ because 
the randomization takes place uniformly.
We use the data for $x=N/2$ in the following, and write it as $G_\Delta(t)$.
Starting from a twisted state,
$G_\Delta(t)$ decays exponentially in a wide range of $p$ and $q$, as shown in Fig.~\ref{fig:5}(a)(b).

%%%%%%%%%%%%%%%%
By fitting $G_\Delta(t)$ by an exponential function, 
we get
\begin{equation}
G_\Delta(t)=\exp\left (-\frac{2pq^2}{a^2N^2}t \right )
\label{eq.tau}
\end{equation}
with the coefficient $a\approx 0.11$.
Now we redefine the timescale for randomization $\tau_{\rm rand}$
by $G_\Delta(\tau_{\rm rand})=0.1$, which gives
\begin{equation}
\tau_{\rm rand} \approx 0.014 \times \frac{N^2}{pq^2}.
\label{eq.taurand}
\end{equation}

\begin{figure}[htbp]
\includegraphics[width=8.6cm]{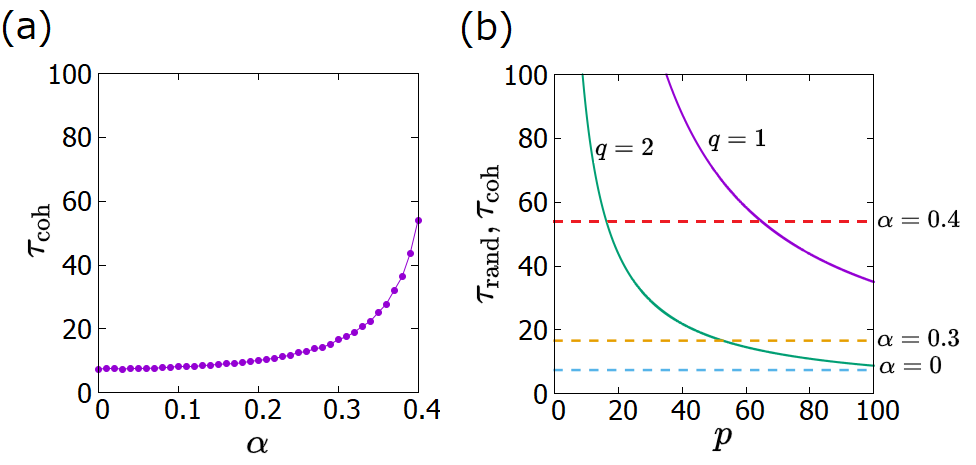}
\caption{(a) The timescale for restoring coherence $\tau_{\rm coh}$ vs. $\alpha$.
(b) Comparison of the timescales.
Solid lines: $\tau_{\rm rand}$ for $q=1,2$. Dashed lines: $\tau_{\rm coh}$ for $\alpha=0,0.3,0.4$.
\label{fig:6}}
\end{figure}

On the other hand, the coupling between oscillators tends to
restore the twisted states.
We turned off the mobility and calculated
the strength of incoherence $\overline{\sigma_\Delta}$
as a function of time, starting from random states;
see Fig.\ref{fig:5}(c) for an attractive coupling
($0 \le \alpha \le 0.4$)
and (d) for a repulsive coupling ($0.8 \le \alpha \le 1$).
In the attractive case,
the decay of $\overline{\sigma}_\Delta$
is slower for larger $\alpha$,
which is interpreted as the result of
the frustration introduced by the phase lag.
We define the timescale for restoring coherence $\tau_{\rm coh}$ 
as the time for $\overline{\sigma_\Delta}$ to decrease from 1 to 0.1.
Fig.\ref{fig:6}(a) shows that $\tau_{\rm coh}$ increases as $\alpha$ is increased from 0 to 0.4.

Now we compare the timescales $\tau_{\rm rand}$
and $\tau_{\rm coh}$ to analyze the competition between
the mobility and coupling.
In Fig.\ref{fig:6}(b), we plot $\tau_{\rm rand}$
given by Eq.(\ref{eq.taurand}) for $q=1, 2$
with solid lines and $\tau_{\rm coh}$ for $\alpha=0, 0.3, 0.4$ with dashed lines.
For $\alpha=0$, we have $\tau_{\rm coh}= 7.4$,
and it is smaller than $\tau_{\rm rand}$ for all $p$ and for $q=1$, $2$.
This is in agreement with the result that
the $q$-twisted states with $q=1$ and $2$ are stable up to
$p=100$ (Fig.\ref{fig:3}(c)).
For $\alpha=0.3$, we get
$\tau_{\rm coh}= 16.6$, which exceeds $\tau_{\rm rand}$
for $q=2$ when $p \gtrapprox 50$.
This suggests that the twisted state with $q=2$ is unstable,
while the twisted state with $q=1$ is still stable for all $p$,
which is also consistent with the full numerical results
(Fig.\ref{fig:3}(c)).
For $\alpha=0.4$ and large $p$, $\tau_{\rm coh}= 54.0$
is smaller than $\tau_{\rm rand}$ even for $q=1$,
which explains the result that
only the synchronized state remains stable (Fig.\ref{fig:4}(a)).
Thus we have shown that the stability of the $q$-twisted states
for $\alpha < \alpha_c(p)$
is determined by the competition between
mobility and coupling, and that the two timescales that
are measured separately can explain the result:
the $q$-twisted state is unstable
if the disturbance caused by the mobility is faster,
and is stable
if the recovery of coherence by the coupling is faster.

%%%%%%%%%%%%%%%%%%%%%%%%%%%%%%%%%%%%%%%%%%%%%%
\subsection{Traveling waves in the transition region}

\begin{figure}[htbp]
\includegraphics[width=8.6cm]{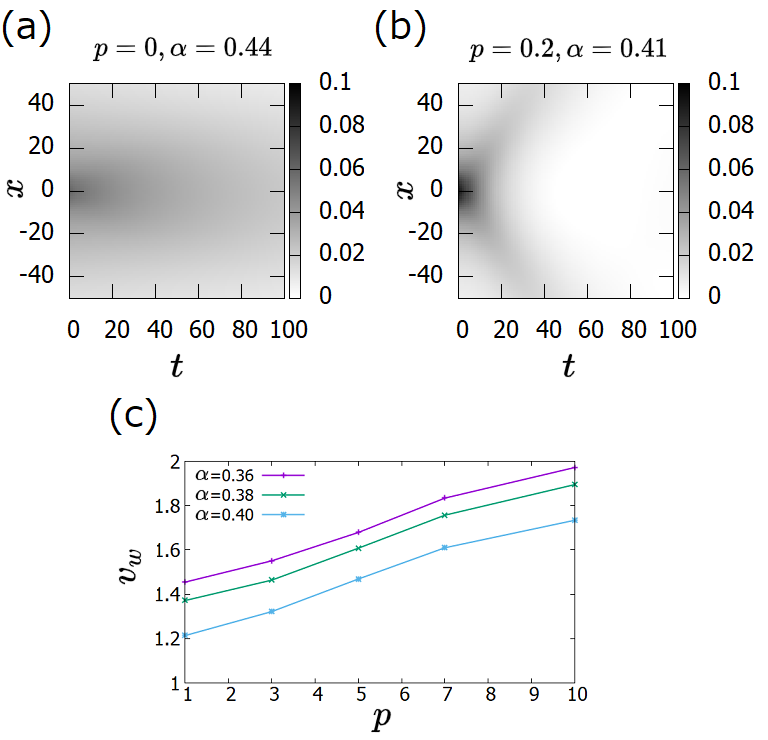}
\caption{The spatiotemporal correlation function
$G_\sigma(x,t)$ for (a) $(\alpha,p)=(0.44,0)$, and
(b) $(\alpha,p)=(0.41,0.2)$,
which corresponds to the patterns in
Fig.\ref{fig:1} (b) and (i), respectively.
The grayscale represents the strength of correlation.
(c) The average speed of the traveling wave $v_w$ versus $p$.
\label{fig:7}}
\end{figure}

We next consider the effect of mobility in the transition region
just above $\alpha = \alpha_c(p)$,
where the strength of incoherence $\overline{\sigma_\Delta}$ is between $0$ and $1$.
The spatiotemporal patterns obtained in this range
is characterized
by the spatiotemporal correlation function of $\sigma_\Delta$:
\begin{equation}
G_\sigma(x,t)=\left \langle \sigma_\Delta(x',t')\sigma_\Delta(x'+x,t'+t)
-
\sigma_\Delta^2(x',t') \right \rangle
_{x',t'}
\label{eq.corre}
\end{equation}
In Fig.~\ref{fig:7}, we show this function
computed in the time window $500 < t' < 1000$
and averaged over 100 independent samples.

%%%%%%%%%%%%%%%%

For $p=0$, the randomly branching pattern
of incoherent states (Fig.~\ref{fig:1}(b))
gives the monotonically decaying
correlation function in Fig.~\ref{fig:7}(a).
For $p>0$, the mesh-like pattern consisting of
intersecting traveling waves (Fig.~\ref{fig:1}(i)) gives
the V-shaped correlation pattern in Fig.~\ref{fig:7}(b).
In order to measure the speed of the traveling waves,
we identify
the average trajectory $x_w(t)$ of each traveling wave
with the ``ridge'' of the correlation function,
where $G_\sigma(x,t)$ is maximal for each $t$.
The average speed $v_w$ of the traveling wave is
measured by fitting $x_w(t)$ with a line.
The dependence of $v_w$ on the mobility $p$
is shown in Fig~\ref{fig:7}(c).
The speed linearly increases with $p$, and slightly decreases
as $\alpha$ is increased above $\alpha_c$.

\begin{figure}[htbp]
\includegraphics[width=6cm]{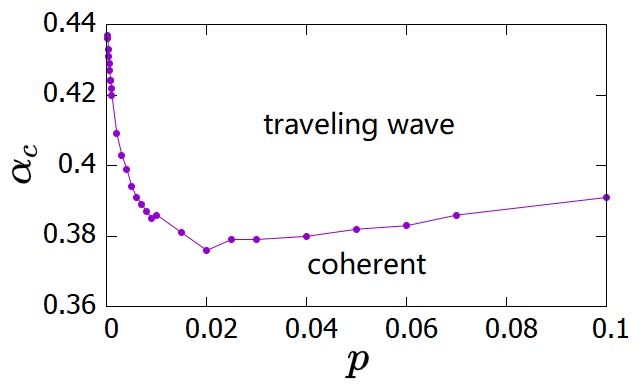}
\caption{The transition threshold $\alpha_c(p)$. For $\alpha < \alpha_c$, only the coherent states remain in the long-time limit, while traveling waves appear for $\alpha > \alpha_c$.\label{fig:8}}
\end{figure}

We define the threshold $\alpha_c$ as the smallest value of
$\alpha$ for which $\overline{\sigma_\Delta}$ takes a nonzero value.
For $p=0$, $\alpha_c$ is known to be about $0.44$. 
We obtained the threshold $\alpha_c(p)$ in Fig.~\ref{fig:8}
by seeing if the system enters the coherent state
in a very long time ($5 \times 10^5$ time units).
For each value of $\alpha$, $\overline{\sigma_\Delta}$ was calculated
for 10 independent samples.
If the system reaches the coherent state
($\overline{\sigma_\Delta}$=0) in every sample,
we judge that $\alpha < \alpha_c$.
If any one of the samples shows a nonvanishing value of $\overline{\sigma_\Delta}$
at $t=5 \times 10^5$, then $\alpha>\alpha_c$.
Fig.~\ref{fig:8} shows that the $p$-dependence of $\alpha_c$
is non-monotonic. For $0<p<0.02$, increasing $p$ lowers the threshold.
For $p>0.02$, $\alpha_c$ slowly increases with $p$. 
For any $p>0$, traveling waves are obtained just above $\alpha_c$, and are replaced by the chimera state at a larger value of $\alpha$.
Direct transition from the coherent to
chimera states are found only for $p=0$.

%%%%%%%%%%%%%%%%

\subsection{The effect of mobility for repulsive coupling}

We next consider the effect of mobility for a repulsive coupling
($\alpha > 0.5$). In the absence of mobility, 
the spatiotemporal pattern consists of multiple domains of
$q$-twisted states with positive and negative values of $q$,
separated by incoherent strips.
The range of the winding number is determined by a linear stability analysis~\cite{li2021large}, according to which the stable range
of the phase difference is $0.215<|\Delta|<0.323$.
(This corresponds to $54 <|q| <80$ for a single-domain twisted state with $N=500$.)
Because of the large phase difference, it is much easier for the mobility
to disturb the twisted states than in the attractive case.
This explains the result that
the system falls into complete incoherence
except for small mobility ($p\delta t\ll 1$)
as seen in Fig.~\ref{fig:2}.
A typical spatiotemporal pattern in the small mobility region
is shown in Fig.~\ref{fig:1}(h), where
the disturbance effect is seen as noisy spots
on the twisted domains.
The threshold of the transition from the noisy pattern
to complete incoherence can be again explained by
the competition of the two timescales.
The timescale of the decay of $\overline{\sigma_\Delta}(t)$ is
in the order of ten time units, and has only minor
dependence on $\alpha$, as seen from Fig.~\ref{fig:5}(d).
(Note that $\sigma_\Delta$ does not drop to zero in the
steady state because of the coexistence of multiple domains. 
In this case, it would be reasonable to redefine $\tau_{\rm coh}$ by the time $\overline{\sigma_{\Delta}}$ 
decays to 0.5.)
Meanwhile, the time for the system to become completely random
in the absence of coupling follows Eq.(\ref{eq.taurand}).
Here, we replace $N/q$ with $2/\Delta$ and use the value $0.215<|\Delta|<0.323$
to obtain $\tau_{\rm rand} \sim 10^{0}/p$.
Therefore, the two timescales are comparable to each other
if $p$ is less than the order of $10^{-1}$,
which is consistent with the threshold
found in Fig.\ref{fig:2}(b).

%%%%%%%%%%%%%%%%%%%%%%%%%%%%%%%%%%%%%%%%%%%%%%%%%%%%%%%%%%%%%%%%%%%%%%%%%%%%%%%%%%
\section{Discussion and Conclusion}

In this paper, we studied the combined effects of
mobility due to random exchange of oscillators
and frustrated coupling due to the phase lag.
Firstly, for $\alpha<0.5$, we showed that the mobility
enlarges the basins of attraction of
twisted states with small $|q|$, including the fully
synchronized state,
by destabilizing the twisted states with large $|q|$.
This is reflected in the decay of the RMS of $q$
as a function of $p$.
For $\alpha=0$, we obtained the power law $\sigma_q \sim p^{-1/4}$,
which is interpreted as follows.
It is shown in a previous study of nonmobile oscillators~\cite{wiley2006size} that the coupling range $R$ and
the standard deviation of $q$ has the relationship
$\sigma_q\approx0.19\sqrt{N/R}-0.11$,
which reduces to $\sigma_q \sim R^{-1/2}$ for $R \ll N$.
On the other hand, mobility effectively extends
the coupling range.
It is derived in Ref.~\cite{uriu2013dynamics}
that $R_{\rm eff} \sim p^{1/2}$ for large $p$,
which means that the effective coupling range $R_{\rm eff}$
is given
by the mean square displacement of each oscillator
in a given microscopic time.
Combining these two results,
we obtain the power law $\sigma_q \sim p^{-1/4}$.

Secondly, we analyzed the timescales
for randomization and restoring coherence
by considering the mobility and coupling separately.
The threshold mobility for destabilizing the $q$-twisted states
is obtained by comparing the two timescales.
For the attractive case, destabilization of a $q$-twisted states
results in hopping to a basin of attraction of another $q$-twisted states
with a smaller $|q|$, or the fully synchronized state.
On the other hand, for the repulsive case,
the coexistence of the $q$-twisted states with large $|q|$ is destroyed by
small mobility, resulting in a fully incoherent state.
Thus mobility has opposite effects for the attractive and repulsive couplings.

Thirdly, we found a mesh-like pattern of traveling waves
for the attractive coupling at the onset of incoherence.
We suggest that local defects in the phase pattern
caused by the exchange of oscillators
trigger the traveling waves.
For $p$ close to 0,
increasing $p$ makes the
traveling waves more easily generated,
and hence the threshold $\alpha_c$ is lowered.
For very large values of $p$, 
all the twisted states will be unstable, 
and only the synchronous state survives.
In this case, the phase pattern is unaffected by the exchange of the oscillators,
and no traveling waves can be generated.
However, the precise mechanism of their formation
is not clear and is left for future work.

\begin{acknowledgments}
This work was supported by JST, the establishment of university fellowships towards the creation of science technology innovation, Grant Number JPMJFS2102 to B. L., and by KAKENHI Grant Number JP21K03396 to N. U.
\end{acknowledgments}
    
%\bibliography{reference}   
\bibliography{li_paper2_bib}

\end{document}